\documentclass[a4paper,11pt,openany]{article}
\usepackage{graphicx}
\usepackage[applemac]{inputenc}
\usepackage{t1enc}
\usepackage{lscape}
\usepackage{amsmath}
\usepackage{amssymb}
\usepackage{cite}
\usepackage{epsfig}
\usepackage{ulem}

\usepackage{xcolor}
\usepackage[colorlinks=true,linkcolor=blue]{hyperref}

\begin{document}

\begin{center}
\Large{\bf Reply to ``Comment on `Validity of path thermodynamic description of reactive systems: Microscopic simulations' ''}
\end{center}

\begin{center}
\large{F. Baras$^{\, a}$, A. L. Garcia$^{\, b}$, and M. Malek Mansour$^{\, c}$}
\end{center}

\begin{center}
\small{ (a) Laboratoire Interdisciplinaire Carnot de Bourgogne, $\quad$\\
 UMR 6303 CNRS-Universit\'e Bourgogne Franche-Comt\'e,\\
 9 Avenue A. Savary, BP 47 870,\\
 F-21078 Dijon Cedex, France

(b) Dept. Physics and Astronomy,\\
 San Jose State University, \\
San Jose, California, 95192 USA

 (c) Universit\'e Libre de Bruxelles CP 231, Campus Plaine,\\
B-1050 Brussels, Belgium}
\end{center}

\begin{abstract}

The Comment's author argues that a correct description of reactive systems should incorporate the explicit interaction with reservoirs, leading to a unified system-reservoirs entity.
However, this proposition has two major flaws.
Firstly, as we will emphasize, this entity inherently follows a thermodynamic equilibrium distribution.
In the Comment, no indication is provided on how to maintain such a system-reservoirs entity in a non-equilibrium state.  Secondly, contrary to the author's claim, the inclusion of system-reservoir interaction in traditional stochastic modeling of reactive systems does not automatically alter the limited applicability of path thermodynamics to problematic reactive systems.
We will provide a simple demonstration to illustrate that certain elementary reactions may not involve %{\color{magenta} (induce ?)} 
any changes in reservoir components, which seems to have been overlooked by the author.

\end{abstract}

\section{Introduction}

The argument presented in the Comment article is based on two separate assertions.
Firstly, the article states: {\it ``Let us further remark that several Markov jump processes may be considered for a given reaction network. This key point is well known''}.
To the best of our knowledge, this statement is likely known only by the author himself, as he introduced it recently in his previous Comment article \cite{Gaspard:2021}.
Furthermore, it contradicts a fundamental principle of probability theory, that is: ``the probability associated with a random event is unique'' (see for example \cite{VanKampen:1983} or \cite{Gardiner:2009}).
We rigorously proved this result in the Introduction of \cite{Baras:2023}.
Recall that the proof relies on the choice of $\mathbb{Z}^n$ as the state space for a homogeneous, isothermal reactive system with $n$ components ($\mathbb{Z}$ represents the set of non-negative integers).
This choice aligns precisely with that of all authors dealing with the stochastic modeling of reactive systems because of its unique correspondence with experimentally measurable quantities \cite{VanKampen:1983,Gardiner:2009}.

Recently, we demonstrated that the validity of path thermodynamics is limited to reactive systems that involve only one elementary reaction leading to each type of observed composition change. \cite{Malek:2017, Malek:2020}.
This proof relies on the traditional stochastic modeling of reactive systems established over half a century ago \cite{VanKampen:1983, Gardiner:2009}.
In order to restore the validity of path thermodynamics in problematic reactive systems, the Comment's author recommended the use of an ``expanded state space'' by incorporating a set of new variables \cite{Gaspard:2021}.
These variables were intended to differentiate the elementary reactions that lead to the same change in composition.

However, as highlighted in our work \cite{Baras:2023}, these newly introduced variables do not correspond to any observable quantities in real-life systems.
This observation served as the primary motivation behind our decision to perform microscopic simulations of reactive systems.
The results of these simulations unequivocally contradict the author's assertion, thereby validating the theoretical predictions based on the traditional modeling of reactive systems \cite{Baras:2023}.
Now it appears that the author has revised his opinion, as there is no mention of the concept of ``expanded state space'' in the present Comment.   Instead, he presents a different approach that we will now address.

In his Comment the author acknowledges the validity of our microscopic simulation results but argues that they fail to account for potential variations in other chemical components that act as control parameters (reservoir quantities). In other words, he contends that the investigation of the statistical properties of reactive systems must explicitly incorporate the interaction between the system and its reservoirs.
In order to illustrate his arguments, the Comment's author considered the same reactive system that we used in our microscopic simulation, that is:
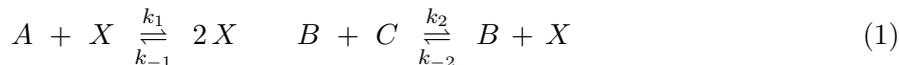
\begin{equation}
\label{Eq1}
A \,\, + \,\, X \,\,\, \mathop{\rightleftharpoons}^{k_1}_{k_{- 1}} \, \,\, 2 \, X \quad  \quad B \,\, + \,\, C \,\,\,   \mathop{\rightleftharpoons}^{k_2}_{k_{- 2}} \, \,\, B \, + \, X
\end{equation}
%where the mole fractions of the reactants $A$, $B$ and $C$ are supposed to remain constant.
where we utilized a well-established procedure to maintain a physico-chemical system out of equilibrium. This procedure involves the system interacting with external reservoirs assumed to be infinitely large, thereby ensuring that their state remains rigorously constant over time.
Author claims in his Comment that, instead of solely considering the variable $X(t)$, while keeping $A, B$, and $C$ constant, we should have analyzed the joint statistical trajectories of $\{X(t), A(t), B(t), C(t)\}$ which takes into account the simultaneous variations of all variables over time (c.f. the last sentence of the fourth paragraph of the Comment).

However, as highlighted in the appendix of our paper \cite{Baras:2023}, the total number of particles in the system-reservoirs entity remains constant, indicating that the state of such an entity is not affected by any external constraint.
Specifically, we wrote,
\begin{quote}
    Finally, note that for both reaction models (2) and (4) the number of $A$, $B$, and $C$ particles and the sum of $X$ and solvent particles $X(t) + S(t)$ remain constant. As such, knowledge of $X(t)$  determines entirely the state of the system at each instant of time.
\end{quote}
Consequently, it can be easily demonstrated that the resulting stationary probability distribution follows a multinomial distribution, which corresponds to a thermodynamic equilibrium distribution.
No indication is provided in the Comment on how to proceed to maintain the system-reservoirs entity in a non-equilibrium state. Not addressing this fundamental issue undermines the arguments criticizing our work.

But there exist a more fundamental objection against Comment author's proposition of the new type of modelling for reactive systems.
Contrary to his claim,  the incorporation of system-reservoir interaction in traditional stochastic modeling of reactive systems does not necessarily alter the limited applicability of path thermodynamics to reactive systems with only one elementary reaction leading to observable compositional changes \cite{Malek:2017,Malek:2020,Baras:2023}.
In fact, certain elementary reactions may simply not involve any changes in reservoir components, a possibility that the author seems to have overlooked.
Consider for example the following set of elementary reactions:
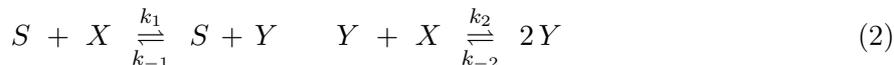
\begin{equation}
\label{Eq2}
S \,\, + \,\, X \,\,\, \mathop{\rightleftharpoons}^{k_1}_{k_{- 1}} \, \,\, S \, + \, Y  \quad  \quad Y \,\, + \,\, X \,\,\,   \mathop{\rightleftharpoons}^{k_2}_{k_{- 2}} \, \,\, 2 \, Y
\end{equation}
both leading either to the change of composition $X, Y \rightarrow X - 1, Y + 1$ (forward) or $X, Y \rightarrow X + 1, Y - 1$ (backward).
%Regardless of how we treat the reservoirs, the path thermodynamic properties of a reactive system involving the above reactions will  necessarily be in contradiction with the corresponding thermodynamic properties \cite{Kondepudi:2008}.
%\Malek{I somehow expand the last paragraph in order to justify the reference to Dilip (below in "magenta").  The reason is that he might be choosen as a referee (as previously). But off course we may leave things the way they are.  Nothing really important}
Regardless of how we treat the reservoirs, the state trajectory of a reactive system involving the set of reactions (\ref{Eq2}) does not incorporate any information that allows us to differentiate them from each other. However, we know from the basic principles of irreversible thermodynamics that the entropy production of a reactive system is the sum of the entropy production associated with each individual reaction \cite{Kondepudi:2008}. Consequently, properties of such a reactive system as given by path thermodynamics will inevitably contradict the actual thermodynamic properties of the system.

In conclusion, we would like to make one final remark. It is interesting to note that the author previously employed the same methodology on multiple occasions, which he now rejects in his Comment.
This includes his seminal 2004 paper, where he developed the path thermodynamic theory of reactive systems \cite{Gaspard:2004}.
Interestingly, in that paper, the author specifically considered the Schl\"{o}gl model (Section IV in \cite{Gaspard:2004}) as an illustrative example of the theory.
It is worth mentioning that the Schl\"{o}gl model is the same type of model we used for microscopic simulation in our article, for which the author now questions the validity.
In the Comment the author appears to contradict statements made in his previous work~\cite{Gaspard:2004}.
In a way, this newest Comment underscores the strength of our earlier works.

\section*{Acknowledgments}

One author (AG) acknowledges support by the U.S. Department of Energy, Office of Science, Office of Advanced Scientific Computing Research, Applied Mathematics Program under contract No. DE-AC02-05CH11231.


\begin{thebibliography}{99}

\bibitem{Gaspard:2021} P. Gaspard, Phys. Rev E {\bf 103}, 016101 (2021).

\bibitem{Baras:2023} F. Baras, A. L. Garcia, and M. Malek Mansour, Phys. Rev. E, {\bf 107}, 014106 (2023).

\bibitem{VanKampen:1983}	N. G. Van Kampen, {\it Stochastic Processes in Physics and
Chemistry}, North-Holland, Amsterdam (1983).

\bibitem{Gardiner:2009}	 C.W. Gardiner, {\it Handbook of Stochastic Methods}, Springer-Verlag (2009).

\bibitem{Malek:2017} M. Malek Mansour and F. Baras, Chaos {\bf 27}, 104609 (2017).

\bibitem{Malek:2020} M. Malek Mansour and A. L. Garcia, Phys. Rev E, {\bf 101}, 052135 (2020).

\bibitem{Gaspard:2004} P. Gaspard, J. Chem. Phys. {\bf 120}, 8898 (2004).

\bibitem{Kondepudi:2008} D. Kondepudi, {\it Introduction to Modern Thermodynamics}, Wiley (2008).


\end{thebibliography}
\end{document}